# A compact heat transfer model based on an enhanced Fourier law for analysis of frequency-domain thermoreflectance experiments

Ashok T. Ramu[1†] and John E. Bowers[1]
[1] Department of Electrical and Computer Engineering
University of California, Santa Barbara
Santa Barbara, CA-93106, USA
† Corresponding author: ashok.ramu@gmail.com

Abstract

A recently developed enhanced Fourier law is applied to the problem of extracting thermal properties of materials from frequency-domain thermoreflectance (FDTR) experiments. The heat transfer model comprises contributions from two phonon channels; one a high-heat-capacity diffuse channel consisting of phonons of mean free path (MFP) less than a threshold value, and the other a low-heat-capacity channel consisting of phonons with MFP higher than this value that travel quasi-ballistically over length scales of interest. The diffuse channel is treated using the Fourier law, while the quasi-ballistic channel is analyzed using a second-order spherical harmonic expansion of the phonon distribution function. A recent analysis of FDTR experimental data suggested the use of FDTR in deriving large portions of the MFP accumulation function; however, it is shown here that the data can adequately be explained using our minimum-parameter model, thus highlighting an important limitation of FDTR experiments in exploring the accumulation function of bulk matter.

## Introduction

The thermoreflectance class of experiments[1] has contributed immensely to our understanding of heat transport on length-scales on the order of the mean-free path (MFP) of phonons, the primary carriers of heat in semiconductor crystals[2]-[8]. The mean-free path accumulation function (MFPAF) introduced by Dames and Chen [9] is a powerful tool in studying ballistic phonon transport. The MFPAF at a given mean-free path Λ is defined as the effective thermal conductivity (ETC) contributed by all phonons with mean-free paths less than or equal to Λ. The MFPAF explains within a unified framework[10] diverse experiments, like the transient gratings[11], time-domain thermoreflectance (TDTR) and frequency domain thermoreflectance (FDTR) experiments[4][5], that probe heat transport on length scales comparable to the phonon mean-free path.

Considering the utility of the MFPAF in explaining length-dependent conductivity measurements in crystalline[4], amorphous[12] and nanostructured materials [13-15] and at thermal interfaces[6], it comes as no surprise that its experimental extraction is the subject of intense research. Recently it has been suggested by Regner *et al.* [4] that a large portion of the MFPAF of crystalline materials (e.g. between 0.4 and 0.95 times the bulk thermal conductivity of crystalline Si at room temperature) may be deduced from FDTR measurements covering modulation frequencies ranging from 200 kHz to 200 MHz. In this work, we demonstrate that the experimental data of [4] can be well explained by a compact model consisting of only two phonon channels, as opposed to an entire spectrum of MFPs. This supports our recent demonstration [16] of a fundamental arbitrariness in the procedure for MFPAF extraction from FDTR experiments propounded by Koh and Cahill [2].

We first describe the theory behind our model, the enhanced Fourier law (EFL), and adapt the equation to the conditions of the FDTR experiment. To this end, we solve the EFL in axisymmetric cylindrical coordinates, and explain the boundary conditions applied in the solution. We then compare the predictions of this model to FDTR experiments and show that excellent fit may be obtained using a simple two-channel approach with only a single MFP for the quasi-ballistic modes, thus contraindicating the use of FDTR experiments for determining the MFPAF.

## The Theory

We apply the enhanced Fourier law (EFL) developed by one of the authors[17] to the FDTR experiment after generalizing it to three spatial dimensions. This model is eminently suited to the problem at hand because: (a) It is closely related to the Fourier law, it being a differential formulation of non-Fourier heat transport consisting of both Fourier law terms and higher order correction terms. Thus most of the mathematical apparatus built for solving Fourier law in various coordinate systems can be applied with a few modifications, as shown in this paper for the cylindrical axisymmetric system. (b) Rigorous derivation[17] from the Boltzmann transport equation (BTE) gives physical meaning to various parameters, instead of treating them as phenomenological fitting parameters, and (c) The EFL is formulated in terms of the mean-free path of low-frequency modes.

The basis of this model of thermal transport is the "two-fluid" assumption[18]. Here the phonon spectrum is divided into two parts- one a high-heat-capacity, high-frequency (HF) part that is in quasi-thermal equilibrium with a well-defined local temperature, and the other, a low-frequency (LF), low-heat-capacity part that is farther out of equilibrium. The LF modes do not interact with each other due to the small phase-space for such scattering[19], but can exchange energy with the HF modes. The HF part of the phonon spectrum corresponds to zone-boundary phonons, which frequently undergo Umklapp scattering. The LF mode phonons, due to their small wave-vectors, chiefly participate in momentum conserving three-phonon processes involving two other HF-mode phonons [19].

The theory of the one-dimensional EFL[17] is based on spherical harmonic expansions of the phonon distribution functions, wherein the high-frequency mode distribution function is truncated at the first order in the expansion, while the low-frequency mode distribution function, which is farther out of thermal equilibrium, is truncated at the second order. The EFL, which may be viewed as a non-local refinement of the Fourier law[20], is extended to three dimensions by noting that under the isotropic assumption, the heat-flux (a) must be composed exclusively of coordinate-invariant components, and (b) must reduce to the one-dimensional Fourier law when the temperature gradient is uniform in any direction. Ignoring the rotational (solenoidal) component of the heat-flux [26], we arrive at the following set of equations, where symbols in bold font represent vectors, while those in normal font their magnitudes or other scalars. For rigorous derivation of these equations from the Boltzmann transport equation, the interested reader is referred to [26].

$$\nabla \cdot \boldsymbol{q} = -C_v \frac{\partial T}{\partial t} + S^{HF}(\boldsymbol{r}, t) \qquad (1)$$

$$\boldsymbol{q} = \frac{3}{5}(\Lambda^{LF})^2 \nabla(\nabla \cdot \boldsymbol{q}) + \frac{3}{5}\kappa^{HF}(\Lambda^{LF})^2 \nabla(\nabla^2 T) - \kappa \nabla T \qquad (2)$$

Here, the net heat-flux $\boldsymbol{q} = \boldsymbol{q}^{LF} + \boldsymbol{q}^{HF}$, $\boldsymbol{q}^{LF}$ = the LF-mode contribution to the heat-flux, $\boldsymbol{q}^{HF}$ = the HF-mode contribution to the heat-flux; $C_v$ is the volumetric heat capacity; $S^{HF}(\boldsymbol{r}, t)$ = spatially and temporally varying external heat source term; $T$ = local temperature of HF modes; $\kappa$ is the net bulk thermal conductivity; $\kappa = \kappa^{LF} + \kappa^{HF}$, $\kappa^{HF}$ = the contribution of HF modes to the bulk thermal conductivity, $\kappa^{LF}$ = the contribution of LF modes to the bulk thermal conductivity; and $\Lambda^{LF}$ = the MFP of the LF modes $= v\tau$ where $v$ is the group-velocity magnitude of all LF modes and $\tau$ = LF mode lifetime. We assume that each and every LF mode has the same lifetime $\tau$, as well as the same group-velocity magnitude $v$, and therefore the same MFP given by $\Lambda^{LF}$. We also state three-dimensional equations for the LF mode and HF mode heat-fluxes $\boldsymbol{q^{HF}}$ and $\boldsymbol{q^{LF}}$ separately:

$$\boldsymbol{q^{LF}} = \frac{3}{5}(\Lambda^{LF})^2 \nabla(\nabla \cdot \boldsymbol{q^{LF}}) - \kappa^{LF}\nabla T \qquad (3)$$

$$\boldsymbol{q^{HF}} = -\kappa^{HF}\nabla T \qquad (4)$$

Fig. 1 shows a schematic of the MFPAF corresponding to our model. Modes with mean-free path less than $\Lambda^{LF}$ are assumed to travel diffusively, thus following the Fourier law. The exact shape of the accumulation function for this region is unknown – it must be stressed that Fig. 1 is schematic. However their thermal properties are aggregated into a single thermal conductivity $\kappa^{HF}$. Since we model all LF modes with the same mean-free path $\Lambda^{LF}$, the remainder of the bulk thermal conductivity, namely $\kappa^{LF} = \kappa - \kappa^{HF}$ is lumped at the mean-free path of $\Lambda^{LF}$, and the MFPAF does not accumulate any further for longer mean-free paths.

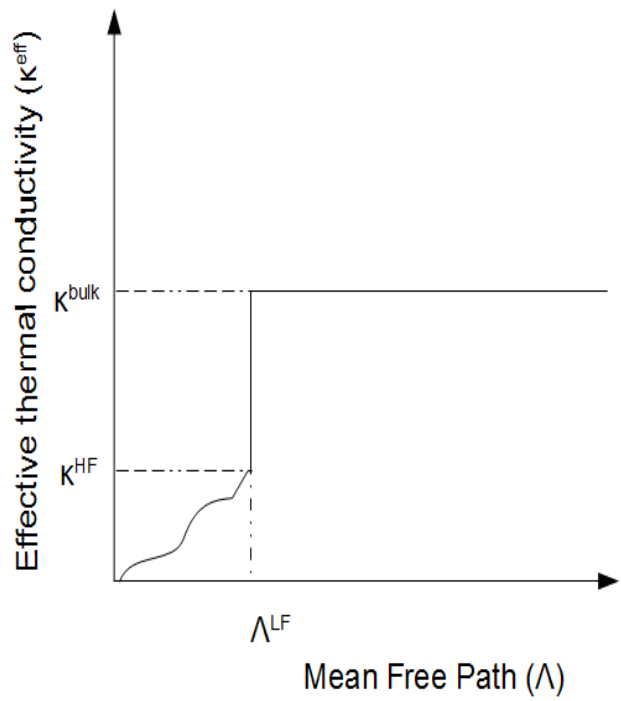

Fig. 1: Schematic of three-parameter mean-free path accumulation function used in the modeling.

## Specialization to the FDTR experiment

In the FDTR experiment of [4], a thin (62 nm) Au/Cr transducer layer is used as the heater (by the application of a pump laser beam) and as a thermometer (by the application of a probe laser beam that monitors surface reflectivity, which is a function of temperature). This transducer is deposited directly on the surface of the substrate to be measured. Combining Eqs. (1) and (2) and transforming to the frequency domain, we arrive at an enhanced heat equation for the substrate, which has no heat generation within ($S^{HF}(\mathbf{r},t) = 0$):

$$-j\omega C_v T = -j\omega \frac{3}{5} C_v (\Lambda^{LF})^2 \nabla^2 T + \frac{3}{5}\kappa^{HF}(\Lambda^{LF})^2 \nabla^2(\nabla^2 T) - \kappa \nabla^2 T \qquad (5)$$

In cylindrical polar coordinates, enormous simplification of the biharmonic term $\nabla^2(\nabla^2 T)$ in this equation is obtained by the introduction of the Hankel transform $\tilde{T}$ of the temperature $T$, defined as

$$\tilde{T}(k,z) = \int_0^\infty T(r,z)\, r J_0(kr) dr \qquad (6)$$

Here $J_0$ represents the ordinary Bessel function of the first kind. Noting crucially that $\nabla^2 \to -k^2 + \frac{\partial^2}{\partial z^2}$ in the Hankel domain, we obtain a fourth-order ordinary differential equation in $z$ for $\tilde{T}$. Solving this by the method of characteristic polynomials, and discarding roots with positive real parts since the substrate is semi-infinite, we obtain a solution of the form

$$\tilde{T} = A(k)e^{p_1 z} + B(k)e^{p_2 z} \qquad (7)$$

where $p_1$ and $p_2$ are the characteristic roots with negative real parts. Coefficients $A$ and $B$ are determined by applying the following boundary conditions: (a) Net heat-flux in the substrate given by the sum of Eq. (3) and Eq. (4) is equal to total heat fluxed by the Au/Cr transducer, and (b) The HF mode heat-flux in the substrate, Eq. (4) is also equal to the total heat fluxed by the Au/Cr transducer. In other words, the transducer does not directly transfer heat to the LF modes of the substrate. Instead, the HF modes are excited first, and then transfer energy to the LF modes through three-phonon processes [21]. This is a reasonable assumption because of the proportionality of scattering rates to the density of final states[22], which is much higher for HF modes than for LF, which in turn follows from the two-fluid assumption stated earlier.

The ordinary Fourier law is used in the 54 nm/9 nm Au/Cr transducer, instead of the isothermal assumption of [4]. For modeling purposes, the Au and Cr materials are combined into one material of effective thickness 62 nm and effective thermal conductivity 110 W/m-K. The boundary between the transducer and the substrate is modeled as an abrupt interface with a boundary conductance $G$ to be extracted from experimental data. The pump and probe laser intensities are modeled as Gaussian distributions with spot sizes given by the $1/e^2$ radii at the beam waist. The solution of the ordinary Fourier law in cylindrical axisymmetric coordinates follows the usual procedure[23], and the surface temperature is extracted as the inverse Hankel transform of $\tilde{T}$ at the top surface of the transducer:

$$T(r,z) = \int_0^\infty \tilde{T}(k,z)\, k J_0(kr) dk \qquad (8)$$

## Results and discussion

Our compact model consists of three material parameters, as shown in Fig. 1: the bulk thermal conductivity, $\kappa$; the contribution of HF modes to the thermal conductivity, $\kappa^{HF}$; and a single LF mode mean-free path $\Lambda^{LF}$. In addition, the boundary thermal conductance $G$ is an experimentally determined parameter. Since the pump laser's $1/e^2$ radius (3.4 microns) is on the same order as the LF-mode phonon mean-free path, the cylindrical axisymmetric geometrical model of this work is strongly indicated, as opposed to one-dimensional models [19][16]. The FDTR experimental data consists of the phase of the surface temperature oscillation plotted as a function of frequency between ~ 200 kHz and 200 MHz. The phase is used because it is relatively insensitive to fluctuations in laser power [8].

Fig. 2 shows the excellent fit obtained with our compact EFL-based model for the FDTR experiment conducted on bulk crystalline Si at 300 K. The parameters are (a) $\kappa$=170 W/m-K, which is reasonable for the thermal conductivity of bulk Si, (b) threshold mean-free path = 1.5 microns, (c) $\kappa^{HF}$=60 W/m-K; this means that only 40% of the heat in silicon is carried by phonons with MFPs below the threshold value, 1.5 microns. The best fit is obtained with $G$=230 MW/m²-K, in rough agreement with [4]. The rest of the parameters, namely the heat capacity of bulk Si and the Fourier law parameters for the transducer are taken from Regner *et al.* [4].

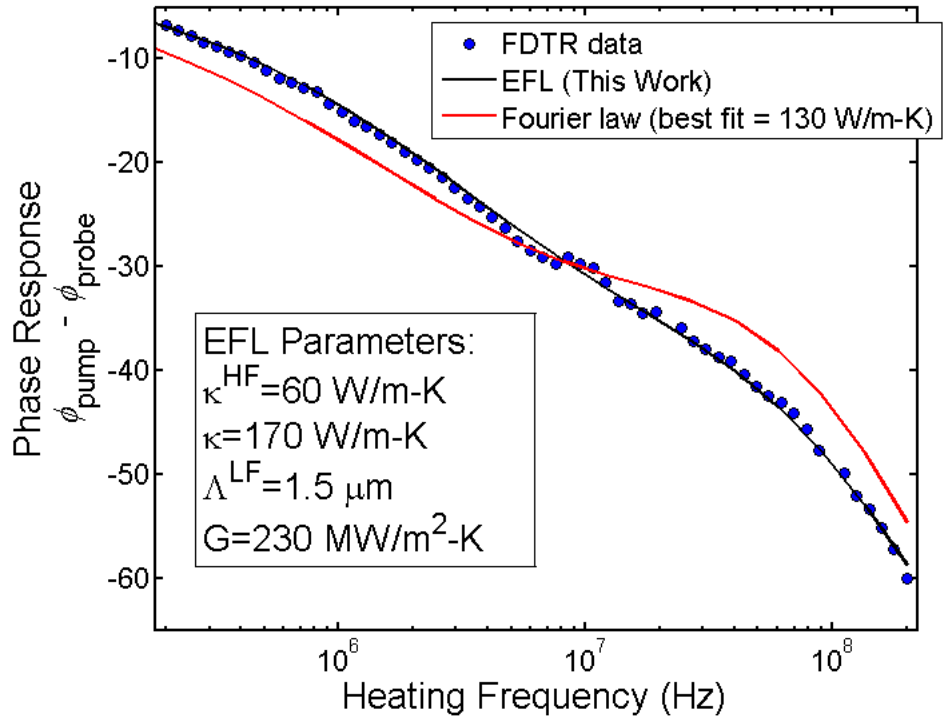

Fig. 2: Phase vs. frequency plot for crystalline Si at 300 K. Experimental data is after Regner *et al.* [4]. The model parameters are shown in the inset (please see Fig. 1); *G* is the interfacial conductance between the transducer and the Si substrate

It seems untenable to extract any more information reliably from this dataset. However, Regner *et al.* have divided the frequency range (60 frequencies evenly spaced on a log scale that spans 200 kHz to 200 MHz) into overlapping windows of 13 points each, and fitted the phase in each window to the Fourier law to extract an effective thermal conductivity (ETC). Each frequency corresponds to a certain thermal penetration depth, given by $\sqrt{\kappa^{eff}/2\omega C_v}$ where $\kappa^{eff}$ is the ETC at that frequency $\omega$ and $C_v$ is the volumetric heat capacity. They assume that phonons with MFP greater than the thermal penetration depth (TPD) travel ballistically across the thermalized region and are therefore lost to the experiment. Thus the TPD at each frequency is used as the cut-off MFP. By plotting the ETC vs. cut-off MFP, they determine the accumulation function. Thus it may be seen that Regner *et al.* effectively use 48 fitting parameters, in addition to the boundary thermal conductance. This is in sharp contrast to our procedure, which uses merely 4 fitting parameters, displayed in Figs. 2 and 3.

As we have noted in [16], one source of arbitrariness in this procedure lies in the abruptness of the TPD cut-off; that phonons go from being entirely diffusive to entirely ballistic as their MFP crosses the TPD. In reality, phonons with MFP of the order of the TPD travel quasi-ballistically, somewhat equilibrating with the thermalized region during their flight. (Completely ballistic phonons do not equilibrate at all, while diffusive phonons equilibrate entirely with the lattice, within the thermalized region that extends to a depth equal to the TPD). Using a different and more realistic criterion for classifying phonons as ballistic or diffusive, we have obtained a drastically different MFP accumulation function from the same data [16]. We note here that an explanation of the results of Regner *et al.* based on just the Fourier law, without invoking ballistic phonon effects, has been offered without experimental proof by [24]. Specifically, it has been assumed that the entire pump laser heat is deposited in the Cr adhesion layer. However, it is well known that significant quasi-ballistic effects occur at micron scales in Si at 300 K[25] and any model that ignores this fact is in our opinion, largely incomplete. Our aim here is to introduce the cylindrical-polar solution of the EFL, and to offer therefrom an alternate and minimalistic interpretation of the data that is in keeping with the spirit of Regner *et al.*

At 400 K however, the fit is not as good, with the best-fit parameters shown in Fig. 3. The larger deviation at high frequencies, where the thermal penetration depth, and therefore the MFP of dominant phonon modes is smaller, suggests the need for a second LF channel of smaller MFP to explain the results. In both cases, 300 K and 400 K, the Fourier law is seen to be quite inadequate in explaining the data, in agreement with [4].

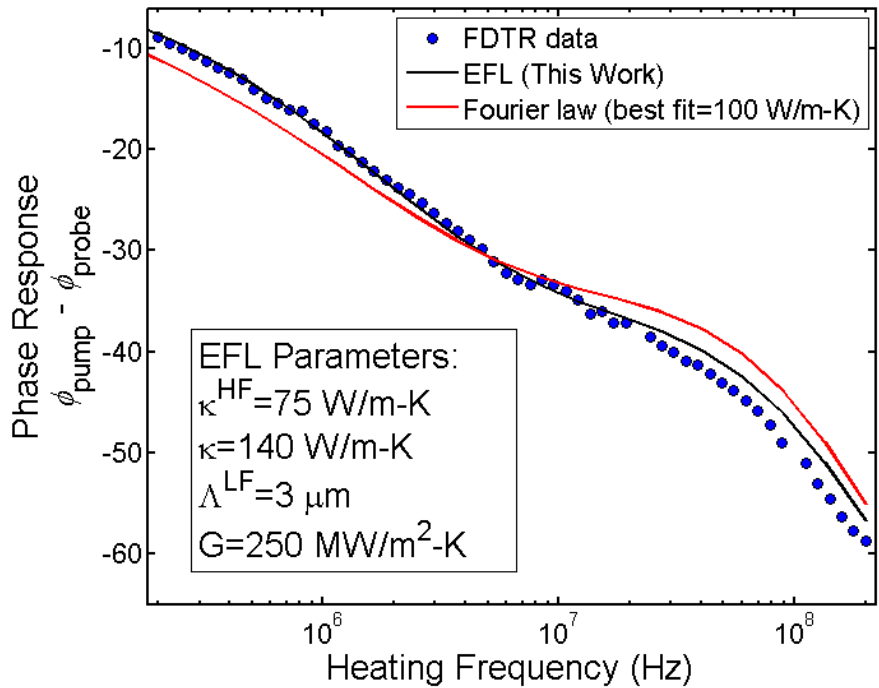

Fig. 3: Phase vs. frequency plot for crystalline Si at 400 K. Experimental data is after Regner *et al.* [4]

In conclusion, we have demonstrated an excellent fit of the phase vs. frequency data from FDTR experiments with the smallest possible parameter set. It is seen that while the Fourier law is too crude to explain the room-temperature data, adding a single MFP channel improves the fit to the point where further improvement is not meaningful; thus no more information may be extracted from the experiment than has been done in this work, contradicting the report of Regner *et al.*[4] who extracted a large portion of the MFPAF from the same dataset.

## Acknowledgments

We wish to thank Dr. Alexei Maznev (Massachusetts Institute of Technology, USA) for proofreading this manuscript, and Prof. Jonathan Malen (Carnegie Mellon University, USA) for kindly providing us with the raw experimental data of Fig. 2 and Fig. 3. This work was funded by the National Science Foundation (NSF), under contract number CMMI-1363207.

References:
1. Paddock, Carolyn A., and Gary L. Eesley, "Transient thermoreflectance from thin metal films", *Journal of Applied Physics* 60, no. 1 (1986): 285-290.
2. Y. K. Koh and D. G. Cahill, "Frequency dependence of the thermal conductivity of semiconductor alloys", *Phys. Rev. B* **76**, 075207, 2007
3. A. J. Minnich, J. A. Johnson, A. J. Schmidt, K. Esfarjani, M. S. Dresselhaus, K. A. Nelson, and G. Chen, "Thermal Conductivity Spectroscopy Technique to Measure Phonon Mean Free Paths", *Phys Rev. Lett* 107, 095901, 2011
4. Keith T. Regner, Daniel P. Sellan, Zonghui Su, Cristina H. Amon, Alan J.H. McGaughey, Jonathan A. Malen, "Broadband phonon mean free path contributions to thermal conductivity measured using frequency domain thermoreflectance", *Nature Comm*. 4, 1640 (2013). See also Supplemental Information.
5. K. T. Regner, S. Majumdar, and J. A. Malen, "Instrumentation of broadband frequency domain thermoreflectance for measuring thermal conductivity accumulation functions", Review of Scientific Instruments 84, 064901 (2013);
6. Ramez Cheaito, John T. Gaskins, Matthew E. Caplan, Brian F. Donovan, Brian M. Foley, Ashutosh Giri, John C. Duda, Chester J. Szwejkowski, Costel Constantin, Harlan J. Brown-Shaklee, Jon F. Ihlefeld, and Patrick E. Hopkins, Thermal boundary conductance accumulation and interfacial phonon transmission: Measurements and theory. Phys. Rev. B 91, 035432 (2015)
7. A. J. Schmidt, X. Chen, and G. Chen, Pulse accumulation, radial heat conduction, and anisotropic thermal conductivity in pump-probe transient thermoreflectance, Rev. Sci. Instrum. 79, 114902 (2008)
8. N. Taketoshi, T. Baba, E. Schaub, and A. Ono, Homodyne Detection Technique using spontaneously generated reference signal in picosecond thermoreflectance meaurements, Rev. Sci. Instrum 74, 12, 2003
9. C. Dames, G. Chen, ''Thermal conductivity of nanostructured thermoelectric materials, Thermoelectrics Handbook: Macro to Nano, Chapter 42, CRC Press, ed. D. Rowe, 2005
10. Justin P. Freedman, Jacob H. Leach, Edward A. Preble, Zlatko Sitar, Robert F. Davis and Jonathan A. Malen, Universal phonon mean free path spectra in crystalline semiconductors at high temperature, Sci. Reports 3, 2963 (2013)
11. Maznev, A. A., Jeremy A. Johnson, and Keith A. Nelson. "Onset of nondiffusive phonon transport in transient thermal grating decay." *Physical Review B* 84, no. 19 (2011): 195206.
12. Jason M. Larkin and Alan J. H. McGaughey, Thermal conductivity accumulation in amorphous silica and amorphous silicon, Phys. Rev. B 89, 144303 (2014)
13. F. Yang and C. Dames, Mean free path spectra as a tool to understand thermal conductivity in bulk and nanostructures, Phys. Rev. B 87, 035437 (2013)
14. Hang Zhang, Chengyun Hua, Ding Ding, Austin J. Minnich. Length Dependent Thermal Conductivity Measurements Yield Phonon Mean Free Path Spectra in Nanostructures, arXiv:1410.6233v1 [cond-mat.mes-hall] 2014
15. G. Romano and J. C. Grossman, Multiscale Phonon Conduction in Nanostructured Materials Predicted by Bulk Thermal Conductivity Accumulation Function, arXiv:1312.7849v3 (2014)
16. A. T. Ramu and J. E. Bowers, Quasi-ballistic phonon transport effects on the determination of the mean-free path accumulation function for the effective thermal conductivity, ArXiv:1502.06005 (2015)
17. A. T. Ramu, "An enhanced Fourier law derivable from the Boltzmann transport equation and a

sample application in determining the mean-free path of nondiffusive phonon modes", J. Appl. Phys. 116, 093501 (2014)
18. B. H. Armstrong, "Two-fluid theory of thermal conductivity of dielectric crystals", *Physical Review B* 23, no. 2 (1981): 883
19. R. B. Wilson, Joseph P. Feser, Gregory T. Hohensee, and David G. Cahill, Two-channel model for nonequilibrium thermal transport in pump-probe experiments. Phys. Rev. B 88, 144305 (2013)
20. Bjorn Vermeersch, Jesus Carrete, Natalio Mingo, Ali Shakouri. Superdiffusive heat conduction in semiconductor alloys -- I. Theoretical foundations. arXiv:1406.7341v2
21. J. M. Ziman, "Electrons and phonons: The theory of transport phenomena in solids." © *Oxford University Press*, (1960)
22. H. Kroemer, Quantum Mechanics for Engineering, Materials Science and Applied Physics (Prentice-Hall, 1994).
23. Aaron J. Schmidt, Ramez Cheaito, and Matteo Chiesa, A frequency-domain thermoreflectance method for the characterization of thermal properties. Rev. Sci. Instrum. **80**, 094901 (2009)
24. R. B. Wilson and D. G. Cahill, "Anisotropic failure of Fourier theory in time-domain thermoreflectance experiments" Nature Comms. 5, 5075 (2014). See supplemental information
25. Jeremy A. Johnson, A. A. Maznev, John Cuffe, Jeffrey K. Eliason, Austin J. Minnich, Timothy Kehoe, Clivia M. Sotomayor Torres, Gang Chen, and Keith A. Nelson, "Direct Measurement of Room-Temperature Nondiffusive Thermal Transport Over Micron Distances in a Silicon Membrane" Phys. Rev. Lett. 110, 025901
26. Ashok T. Ramu and John E. Bowers, "On the solenoidal heat-flux in quasi-ballistic thermal conduction", arXiv preprints, arXiv:1505.02465v1, (2015)